\documentclass[amsmath,amssymb,aps,prl,reprint,superscriptaddress,footnoteinbib]{revtex4-1}
\usepackage{amsmath,amssymb}
\usepackage[english]{babel}
\usepackage{ bbold }
\usepackage{upgreek}
\usepackage{dsfont}
\usepackage{graphicx}
\usepackage{comment}
\usepackage{braket}
\usepackage{colortbl}
\usepackage{blindtext}
\usepackage[dvipsnames]{xcolor}

\newcommand{\changed}[1]{#1}

\begin{document}

\title{Entangling levitated nanoparticles by coherent scattering}
	
\author{Henning Rudolph}
\affiliation{Faculty of Physics, University of Duisburg-Essen, Lotharstra\ss e 1, 47048 Duisburg, Germany}

\author{Klaus Hornberger}
\affiliation{Faculty of Physics, University of Duisburg-Essen, Lotharstra\ss e 1, 47048 Duisburg, Germany}

\author{Benjamin A. Stickler}
\affiliation{Faculty of Physics, University of Duisburg-Essen, Lotharstra\ss e 1, 47048 Duisburg, Germany}
\affiliation{QOLS, Blackett Laboratory, Imperial College London, London SW7 2BW, United Kingdom}

\begin{abstract}
We show how entanglement between two optically levitated nanoparticles can be generated and detected by coherent scattering of tweezer photons into a single cavity mode. 
Triggered by the detection of a Stokes photon,
the tweezer detuning is switched from the blue to the red;
entanglement 
is then verified by the conditioned anti-Stokes photon flux, which oscillates with the mechanical beat frequency.
The proposed setup is realizable with near-future technology and opens the door to the first experimental observation of non-classical center-of-mass correlations between two or more levitated nanoscale objects.
\end{abstract}

\maketitle

{\it Introduction---} Nanoparticles optically levitated in high vacuum can be accurately controlled by laser light, while staying well isolated from the ambient environment. This makes them ideally suited for high-precision sensing applications \cite{chaste2012,ranjit2016,hempston2017} and for the next generation of macroscopic quantum superposition tests \cite{romeroisart2011a,bateman2014,arndt2014a,wan2016a,maqro2016,stickler2018b},
most of which require cooling into the deep quantum regime. 

Coherent scattering cooling is a promising new approach to prepare a massive nanoparticle in its motional quantum groundstate \changed{\cite{salzburger2009,delic2019,windey2019,uros}}. As in conventional cavity cooling, the particle is levitated by an optical tweezer inside a high-finesse cavity \cite{chang2010,romeroisart2010,barker2010,asenbaum2013,kiesel2013,millen2015,fonseca2016}. However, if the tweezer is slightly red-detuned with respect to the cavity resonance, the nanoparticle coherently scatters tweezer photons into the cavity mode and thereby efficiently reduces its motional energy \cite{gonzalesballestero2019}. In this article we show that extending this setup to two particles interacting with the same cavity mode provides an attractive platform for the first observation of center-of-mass entanglement between two or more levitated nanoscale objects.

Entanglement was recently observed between two spatially separated clamped micromechanical oscillators \cite{riedinger2018,ockeloen2018}, an impressive technological feat and an important step towards future applications of non-classical correlations in optomechanics-based quantum technologies \cite{aspelmeyer2014}. 
Even if the capacity of such experiments 
to probe macrorealist extensions of quantum mechanics 
may be limited \cite{bassi2013,schrinski2019},
they bring into focus the counter-intuitive nature of quantum mechanical non-locality \cite{markinovic2018}.
Entanglement between two optically levitated nanoparticles is expected to persist for much longer than between clamped structures, due to the former's exquisite environmental isolation. Generating and detecting quantum correlations between levitated objects is thus a crucial step for future implementations of quantum technology and  non-locality experiments.

Here we present a scheme to generate and read-out entanglement between levitated nanoparticles in a single cavity.  It is based on changing the detuning of the tweezers from the blue to the red sideband conditioned on the detection of a Stokes photon.
Since the method avoids laser phase noise, coherence times on the order of hundreds of micro-seconds can be expected  for realistic experimental parameters.

\begin{figure}
	\centering
	\includegraphics[width=1.05\linewidth]{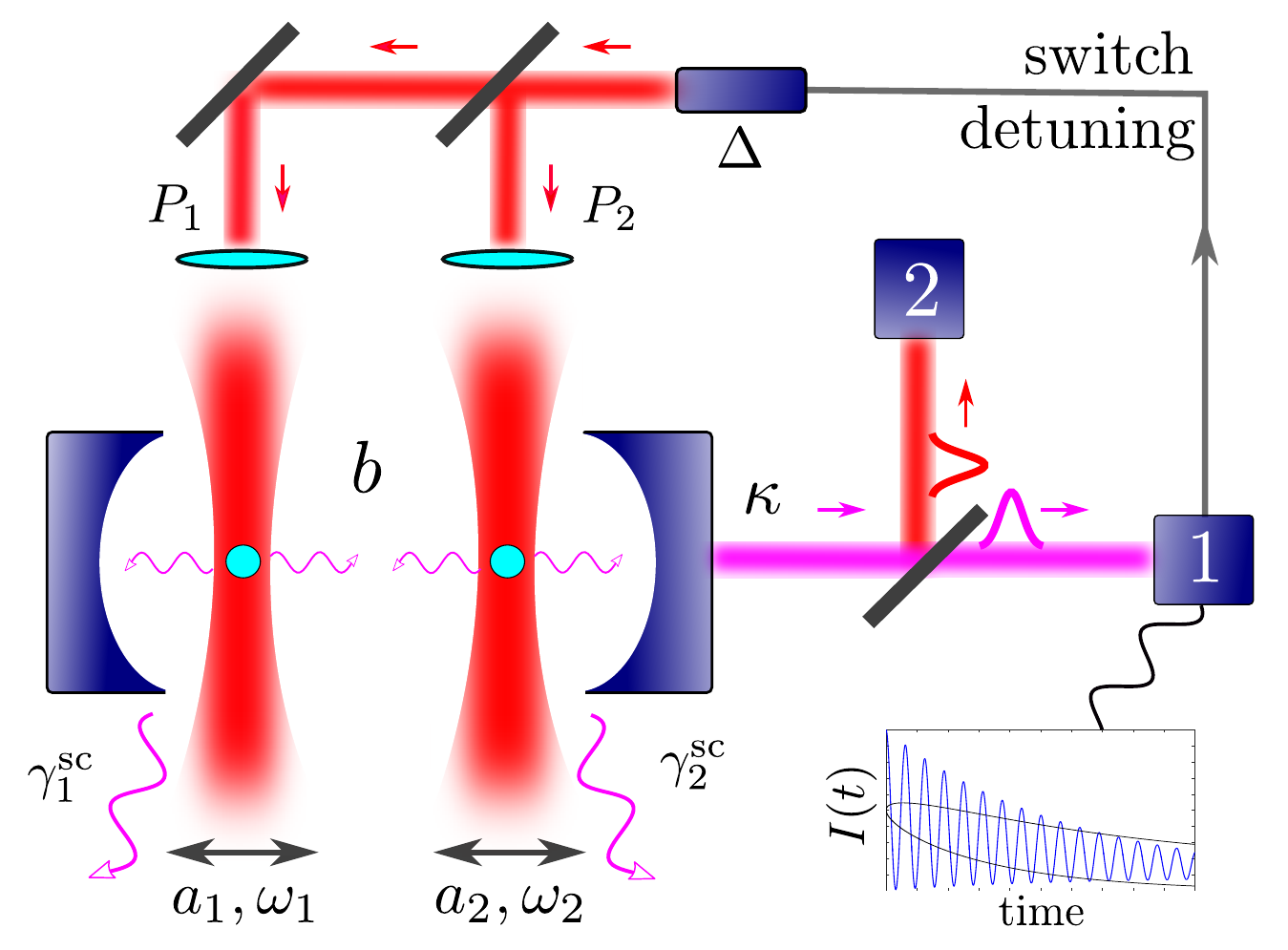}
	\caption{Two nanoparticles, trapped by optical tweezers of powers $P_1$, $P_2$ are weakly coupled to an optical cavity of linewidth $\kappa$. Their harmonic motion, described in transverse direction by the ladder operators $a_{1,2}$ and frequencies $\omega_{1,2}$,  is initially cooled to the groundstate. Setting the tweezer detuning to $\Delta = (\omega_1 + \omega_2)/2$ (blue sideband) a Stokes photon, coherently scattered into the intra-cavity field $b$, may then be detected in the output mode (photon detector 1). This effectively maps the two particles onto an entangled state. Triggered by the Stokes photon detection, the detuning is reversed to $\Delta = -(\omega_1 + \omega_2)/2$ and the appearance of an anti-Stokes photon is measured as a function of time. Entanglement is verified by observing that this flux $I(t)$, oscillating with the mechanical beat frequency, exceeds a time-dependent threshold determined by the Rayleigh scattering rates $\gamma_j^{\rm sc}$. {The beam splitter includes a filter directing off-resonant photons to detector 2, thus removing them from the feedback loop; possible occurrences of a second Stokes photon are thus discarded.}
	} \label{fig:sketch}
\end{figure}

{\it Outline of the proposed experiment---} The envisaged experimental setup is sketched in Fig.~\ref{fig:sketch}. Two nanoparticles are levitated inside a cavity with mode frequency $\omega_{\rm c}/2\pi$, waist $w$, and {field decay rate} $\kappa$. A laser beam detuned by $\Delta = \omega_{\rm L} - \omega_{\rm c}$ is split to drive two tweezers with powers  $P_{1,2}$ and waists $w_{1,2}$, each trapping a single particle. 

Tweezers and cavity are aligned such that the tweezer axes intersect the cavity symmetry axis, without phase lag between the tweezers. If the nanoparticle are placed on a node of the cavity field their dynamics along the cavity axis decouple from their other degrees of freedom and the motion turns effectively harmonic with trapping frequencies $\omega_{1,2}$ and mode operators $a_{1,2}$. These trapping frequencies are slightly detuned by $\delta \omega_{\rm m} = \omega_2 - \omega_1$, by adjusting the tweezer powers or  waists.

Coherent scattering of the tweezer photons \cite{delic2019,windey2019,salzburger2009} \changed{can} cool the particles to the \changed{quantum ground state \cite{uros}}. Once the ground state is reached and the tweezer focus is placed onto a cavity node the cavity field vanishes.

Entanglement between the nanoparticles is generated by tuning at time $t=0$ the laser onto the mean mechanical frequency $\Delta = \overline{\omega} \equiv (\omega_1 + \omega_2)/2$. Subsequent Stokes scattering of a tweezer photon into the cavity mode prepares one of the two nanoparticles in its first excited state. Crucially, detecting this photon in the cavity output does not reveal which particle has been excited and thus effectively prepares them in an entangled state. 

In order to detect the entanglement, the laser detuning is reversed upon detection of the Stokes scattered photon to  $\Delta = - \overline{\omega}$, and the arrival time of the first anti-Stokes scattered photon is recorded by detector 1. ({A filter serves to discard all off-resonant photon detections.})
Repetition of the entire scheme yields the average conditional flux of anti-Stokes scattered photons at time $t$ after the Stokes scattered photon has been detected.

This photon flux is given by
\begin{align}\label{measure}
I(t) = 2\frac{g^2}{\kappa} \langle (a_1 + a_2)^\dagger (a_1 + a_2) \rangle_t,
\end{align}
with $g = (g_1 + g_2)/2$ the mean nanoparticle-cavity coupling rate, as follows from the input-output formalism  for $g/\kappa \ll1$ \cite{borkje2011}. {The photon flux} oscillates with the mechanical detuning and decays with the photon scattering decoherence rates $\gamma_{1,2}^{\rm sc}$. Importantly, one can show that for classically correlated, i.e. separable, mechanical states this current always lies within certain bounds. Observing the flux to surpass these bounds therefore verifies the presence of mechanical entanglement.

{\it Coupled nanoparticle-cavity dynamics---} To quantitatively assess the outlined experimental proposal, we calculate the coupled cavity-particle dynamics in presence of realistic sources of environmental decoherence. The harmonic frequencies of the trapped nanoparticles are determined by their susceptibility $\chi$, their mass density $\varrho$ as well as by the tweezer waists and powers through $\omega_j = 2\sqrt{ \chi P_j} /w_j^2 \sqrt{\pi c \varrho}$. (Here and in what follows we assume that the particles densities and susceptibilities are identical.)

How strong the nanoparticles couple to the cavity mode depends on the mode and particle volumes $V_{\rm c}$ and $V_{1,2} $, and on the wavenumber $k = \omega_{\rm c}/c$. Close to the tweezer focus the cavity-particle coupling is linear in the mechanical amplitude and the corresponding coupling rate takes on the form \cite{gonzalesballestero2019}
\begin{align}
g_j &= \frac{\chi}{2w_j} \sqrt{\frac{P_j k^3 V_j}{\pi V_{\rm c}\varrho \omega_j}}.
\end{align}

In the frame rotating with the laser frequency, the Hamiltonian for the coupled cavity-nanoparticle system assumes the form of three linearly coupled harmonic oscillators,
\begin{align}
\frac{H}{\hbar} =& -\Delta b^\dagger b + \sum_{j=1,2}\omega_j a_j^\dagger a_j - \sum_{j=1,2}g_j( b +  b^\dagger)(a_j + a_j^\dagger),
\end{align}
with the cavity mode operator $b$. Taking cavity loss due to its finite linewidth and nanoparticle decoherence due to scattering of tweezer photons into account, the dynamics of the total state operator $\rho_{\rm tot}$ can be expressed as a Lindblad quantum master equation
\begin{align}\label{mastertot}
\partial_t\rho_{\rm tot} = & - \frac{i}{\hbar} \lbrack H,\rho_{\rm tot} \rbrack + 2\kappa \mathcal{L}[b]\rho_{\rm tot} \nonumber \\
& + \sum_{j = 1,2} \gamma_j^{\rm sc} \left ( \mathcal{L}[a_j]\rho_{\rm tot} + \mathcal{L}[a_j^\dagger]\rho_{\rm tot} \right ).
\end{align}
Here, the superoperator $\cal L$ is defined as ${\cal L}[c]\rho_{\rm tot} = c \rho_{\rm tot}c^\dagger -  c^\dagger c \rho_{\rm tot} /2 - \rho_{\rm tot}c^\dagger c /2$, and for sufficiently low pressures the decoherence rates are dominated by Rayleigh scattering close to the tweezer center \cite{Jain}
\begin{align}
\gamma_j^{\rm sc} &= \frac{P_j \chi^2 V_j k^5}{15 \pi^2 c \varrho \omega_j w_j^2 }.
\end{align}

In the following we consider the case of weak coupling, $g_j \ll \kappa$, and use the fact that the cavity is initially empty. Thus the probability of finding a photon in the cavity is small for all relevant times $t \ll 1/\gamma_j^{\rm sc}$, implying that the photon occupations larger than unity can be neglected. The remaining nonzero density matrix elements {$\rho_{\ell \ell'}=\langle \ell | \rho_{\rm tot}| \ell'\rangle$, with $\ell,\ell' \in \{0,1\}$ cavity photon numbers},  obey a set of four coupled operator-valued differential equations, directly obtained from  \eqref{mastertot}.

By solving these for $\rho_{10}$ and $\rho_{01}$ in the weak coupling approximation to leading order in $g_j /\kappa$ one can derive a closed equation for 
the dynamics of the reduced mechanical state operator $\rho = \rho_{00}+\rho_{11}$ {for times $\kappa t\gg 1$}. Assuming $\delta\omega_{\rm m} \gg g_j^2/\kappa$ all terms rotating either with the mechanical frequency or with the mechanical detuning may be neglected. It then reduces to
\begin{align} \label{eq:redmaster}
\partial_t \rho = & -i \sum_{j=1,2} (\omega_j + \delta \omega_j^{\rm opt}) \lbrack a_j^\dagger a_j, \rho \rbrack \nonumber \\
& + \sum_{j = 1,2} \left (\gamma_j^{-} \mathcal{L}[a_j]\rho + \gamma_j^{+} \mathcal{L}[a_j^\dagger]\rho \right ).
\end{align}
The nanoparticles  behave  effectively as two independent  harmonic oscillators coupled to an environment.

Their mechanical frequencies are shifted by the optical spring effect \cite{aspelmeyer2014}
\begin{align}
\delta\omega_j^{\rm opt} = g_j^2 \left( \frac{\Delta + \omega_j}{\kappa^2 + (\Delta + \omega_j)^2} + \frac{\Delta - \omega_j}{\kappa^2 + (\Delta - \omega_j)^2} \right),
\end{align}
and the mechanical heating and cooling rates %
are given by
\begin{equation}
\gamma_j^{\pm} =  \gamma_j^{\rm sc} + \frac{2 g_j^2 \kappa}{\kappa^2 + (\Delta \mp \omega_j)^2}.
\end{equation}
The latter will eventually lead to  an effective thermalization due to radiation pressure shot noise \cite{marquardt2007} towards a mean phonon occupation of $n_j = \gamma_j^+/\gamma_j$, where $\gamma_j = \gamma_j^- - \gamma_j^+$.

Moreover, {using the same weak-coupling approximation as above,}  one can express $\rho_{11}$ in terms of the reduced mechanical state operator as $\rho_{11}(t)= B \rho(t) B^\dagger$ for times $\kappa t\gg 1$ with
\begin{align}\label{rho11}
B =  ig\left(\frac{a_1 + a_2}{\kappa - \text{i}(\Delta + \overline\omega)} + \frac{ (a_1 + a_2)^\dagger}{\kappa - \text{i}(\Delta - \overline\omega)}\right) \,.
\end{align}
We approximated $g_{1,2}\simeq g$ and $\omega_{1,2}\simeq \overline{\omega}$ for simplicity.

The reduced master equation \eqref{eq:redmaster} yields closed equations for the moments of the mechanical mode operators $a_j$. A direct calculation shows that the mean occupations evolve as
\begin{subequations}\label{solution}
\begin{align}
\langle a_j^\dagger a_j \rangle_t &= \langle a_j^\dagger a_j \rangle_0 \text{e}^{-\gamma_j t} + n_j (1-\text{e}^{-\gamma_j t}).
\end{align}
In a similar fashion, one obtains that the mode coherences oscillate with the effective mechanical detuning $\delta\omega_{\rm eff} = \delta \omega_{\rm m} + \delta \omega_2^{\rm opt} - \delta \omega_1^{\rm opt}$ and decay with the mean damping rate $\gamma = (\gamma_1 + \gamma_2)/2$,
\begin{align}
\langle a_2^\dagger a_1 \rangle_t &= \langle a_2^\dagger a_1 \rangle_0 \text{e}^{i\delta\omega_{\rm eff}t - \gamma t}.
\end{align}
Finally, the population coherences also approach stationary values in a more complicated way,
\begin{align} \label{eq:solutionc}
\langle a_1^\dagger a_1 a_2^\dagger a_2 \rangle_t &= \langle a_1^\dagger a_1 a_2^\dagger a_2 \rangle_0 \text{e}^{-2\gamma t} + n_1 n_2 (1-\text{e}^{-2 \gamma t}) \nonumber \\
&+ n_1 (\langle a_2^\dagger a_2 \rangle_0 - n_2)(\text{e}^{-\gamma_2 t} - \text{e}^{-2  \gamma t}) \nonumber \\
&+ n_2 (\langle a_1^\dagger a_1 \rangle_0 - n_1)(\text{e}^{-\gamma_1 t} - \text{e}^{-2 \gamma t}).
\end{align}
\end{subequations}
The dynamics \eqref{solution} of the moments will be crucial for verifying that entanglement was indeed present.

\begin{figure}
\centering
\includegraphics[width=0.95\linewidth]{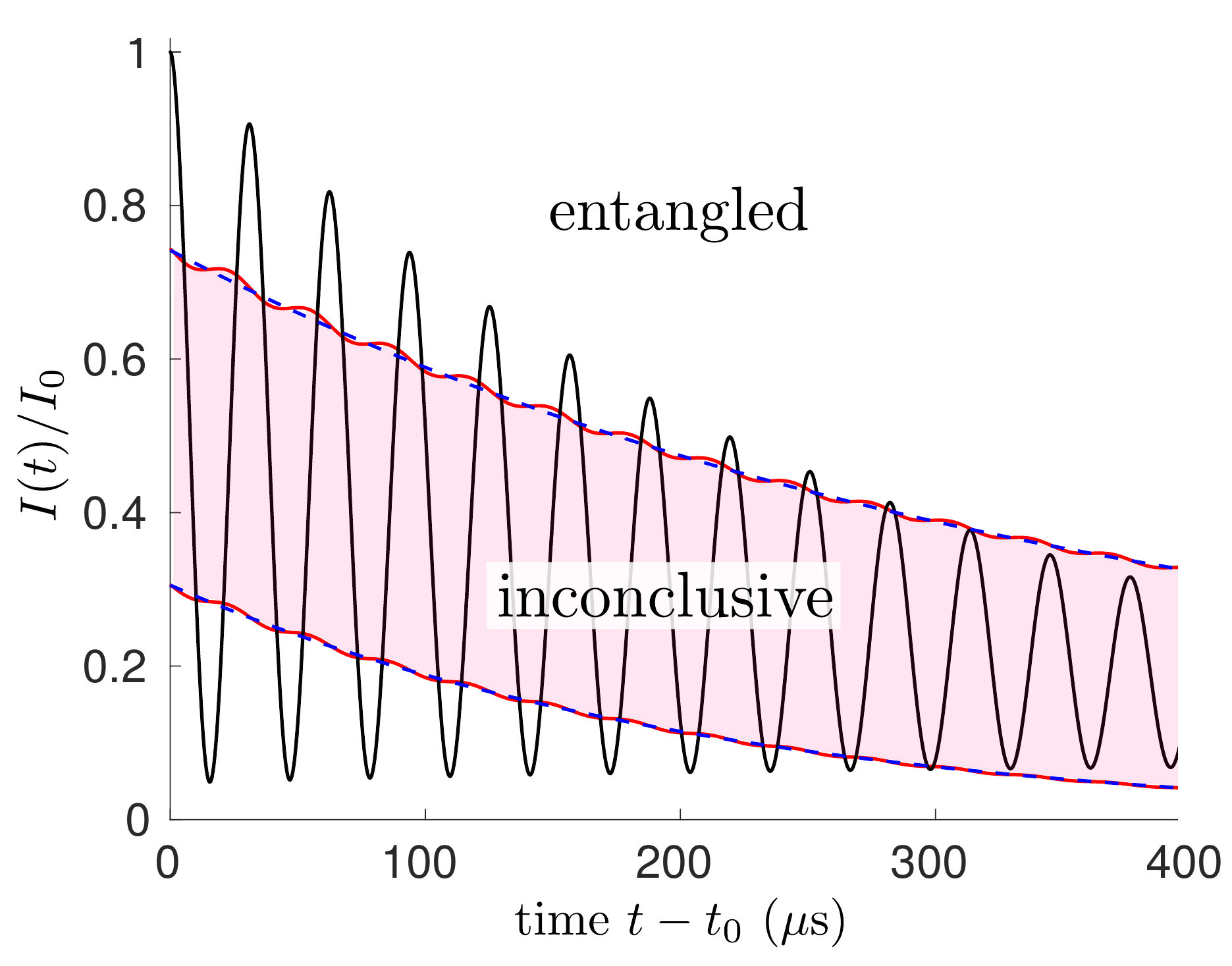}
\caption{Flux of anti-Stokes scattered photons (black solid line) conditioned on the detection of a Stokes scattered photon \changed{for two Si spheres ($\varrho = 2336$\,kg/m$^3$, $\chi = 2.4$) of $10$\,nm radius}, {assuming an initial ground state population of 95\%}. Measuring a photon flux outside the shaded region indicates that entanglement has been generated by the measurement of the Stokes scattered photon. The boundary of this region was calculated numerically from \eqref{mastertot} (red line) and approximated analytically with \eqref{solution} (blue dashed line). The tweezer and cavity parameters are $2 \pi/k = 1560$\,nm, $P_j = 1.5$\,W, $w_j = 720$\,nm, $\ell = 12$\,mm,  $w = 30$\,$\mu$m, $\kappa/2\pi = 318$\,kHz, yielding the \changed{coupling frequencies $g_j = 61$\,kHz}, the trapping frequency $\overline{\omega}/2\pi = 785$\,kHz, and the decoherence rates $\gamma_j^{\rm sc} = 145$\,Hz for a mechanical detuning of $\delta\omega_{\rm m} = 32$\,kHz. The residual gas pressure is assumed to be below $10^{-9}$\,mbar. 
	 } \label{fig:plot}
\end{figure}

{\it Entanglement generation and read-out---} The particles are initially prepared in the mechanical groundstate $|00 \rangle$. To generate entanglement the laser detuning is initially set to $\Delta = \overline \omega$. For times $\gamma_j t_0 \ll 1 \ll \kappa t_0$ the dynamics of the reduced state are essentially determined by the unitary contribution in \eqref{eq:redmaster}. Measuring a Stokes scattered photon at time $t_0$ effectively reduces the mechanical state to ${\rm tr}(b\rho_{\rm tot}(t_0) b^\dagger)_{b}\propto\rho_{11}(t_0)$,  as determined by \eqref{rho11} with $\Delta = \overline \omega$. Here, ${\rm tr}(\cdot)_{b}$ denotes the partial trace over the photonic Hilbert space. One thus obtains, {with an infidelity on the order of $g^2/\kappa^2$,} the entangled state $\rho'(t_0) = |\psi_0\rangle \langle \psi_0|$ with
\begin{equation}\label{psi0}
|\psi_0\rangle = \frac{1}{\sqrt{2}} \left ( |10\rangle +|01\rangle \right ).
\end{equation}

Right after detection of the Stokes scattered photon, the detuning is switched to $\Delta = -\overline \omega$, on the time scale of $1/\kappa$. The reduced state $\rho'$ still evolves according to \eqref{eq:redmaster};
it determines the conditional flux (\ref{measure}) of anti-Stokes scattered photons leaving the cavity  at time $t>t_0$. We note that this flux can also be obtained by neglecting the phonon creators in (\ref{rho11}), i.e.\ the second Stokes scattered photons in $\rho'_{11}(t)=B\rho'(t)B^\dagger$; this directly yields $I(t) =2\kappa \, {\rm tr}(b \rho_{\rm tot}'(t) b^\dagger) =  2 \kappa\, {\rm tr}(\rho'_{11}(t))$. 

The expected flux of anti-Stokes scattered photons, $I(t) \propto \sum_j \langle a_j^\dagger a_j \rangle_t+ 2 {\rm Re}(\langle a_2^\dagger a_1\rangle_t)$, can be calculated from \eqref{solution} by using that $\langle a_{j}^\dagger a_{j}\rangle_{t_0}  =\langle a_{2}^\dagger a_{1}\rangle_{t_0} = 1/2$. It oscillates with the effective mechanical detuning $\delta \omega_{\rm eff}$, while decaying exponentially with the decoherence rates $\gamma_j$. 
The amplitude of these photon flux oscillations can be used to verify the presence of entanglement in the state $\rho'$.

In particular, we use that for classically correlated states \cite{borkje2011}
\begin{align}\label{witness}
|\langle a_2^\dagger a_1 \rangle_t | \leq \sqrt{\langle a_1^\dagger a_1 a_2^\dagger a_2 \rangle_t}
\end{align}
for any pair of operators $a_1$ and $a_2$ acting on different Hilbert spaces. 
Thus, if the amplitude $\vert \langle a_2^\dagger a_1 \rangle_t \vert$ of the photon flux oscillations  violates this inequality, the presence of initial entanglement between the two nanoparticles has been verified.

A genuine entanglement witness would require measuring the occupation correlation $\langle a_1^\dagger a_1 a_2^\dagger a_2 \rangle_t$, which is in general difficult. Since the initial state (\ref{psi0}) implies $\langle a_1^\dagger a_1 a_2^\dagger a_2\rangle_{t_0} = 0$ 
we instead  compare the measured photon flux with the bounds  obtained from \eqref{eq:solutionc} by assuming the initial occupation correlation to vanish  (using $t_0$ as the initial time). 
Observation of a photon flux exceeding this bound verifies entanglement based on the validity of standard quantum theory and the presented description of the experiment, which neglects multi-photon excitations. 

In Fig.~\ref{fig:plot} we show the expected conditional photon flux of the proposed experiment, {assuming an initial ground state population of 95\%}. \changed{One observes that} the photon flux (black solid line) oscillates with the mechanical beat frequency $\delta\omega_{\rm eff}$ \changed{and recurringly exceeds the verification bound (shaded region) over a period of hundreds of microseconds}. The dashed line indicates the analytical estimate for the  bound, as obtained by combining (\ref{solution}) with (\ref{witness}). It compares well with the numerically exact bound (red solid line), obtained by integrating the master equation (\ref{mastertot}) for the total state of particles and cavity field. {The main effect of the assumed finite initial excited state population is to broaden the inconclusive region, while the conditioned photon flux is only weakly affected.}

For longer measurement times decoherence due to Rayleigh scattering of tweezer photons and due to cavity photon shot noise reduces the oscillation amplitude until entanglement is no longer observable. 
The time during which entanglement can be verified {for an initial ground state} follows straightforwardly from \eqref{solution} as
\begin{align}
t_\text{dec} \simeq \frac{1}{\gamma} \ln \left( \frac{2{n}+\sqrt{2}-1}{2 {n}} \right),
\end{align}
assuming  ${n}\equiv(n_1+n_2)/2 \simeq n_j$ and $\gamma_j \simeq \gamma$. 

{We note that imperfections in both the particle positioning and the photon detections do not impair the entanglement verification scheme, as long as the tweezer light is filtered  from detector 1. The main effect of displacing the particles from the cavity nodes, e.g. by 8\,nm \cite{delic2019}, is a drive of the cavity field at the tweezer frequency \cite{gonzalesballestero2019}. \changed{In the scenario considered in Fig.~\ref{fig:plot} this results in the presence of below ten photons;}
the particle positioning can be optimized by monitoring detector 2. 
Photon loss or an imperfect quantum efficiency of detector 1 reduces the photon flux only by a constant factor \cite{wiseman2009}, as can be seen from the fact that the solution of the master equation (\ref{mastertot}), and therefore the expectation value (\ref{measure}), involves an average over all possible (registered or unregistered) photon detection events. 
}

{\it Entangling $N$ particles---} The here presented scheme works even if more than two nanoparticles are levitated inside the cavity. For weak coupling,  $\sum_{j=1}^N g_j\ll \kappa$, the respective generalizations of \eqref{measure} and \eqref{rho11} involve the $N$-particle annihilation operator $A = a_1+\cdots+a_N$ in place of $a_1+a_2$, and in the resulting reduced master equation \eqref{eq:redmaster} the sum runs over $N$ independent damped harmonic oscillators. %
Relations \eqref{solution} are still valid in the multi-particle case, with corresponding rates and frequencies.

Enacting the entanglement generation protocol thus excites a single out of $N$ oscillators, effectively preparing the W-state $\rho' = | \psi_0 \rangle \langle \psi_0| $ with
$|\psi_0\rangle \propto \left ( |1 0 0 \cdots 0\rangle + | 0 1 0 \cdots 0\rangle + \ldots + |0 0 0 \cdots 1 \rangle \right ).
$
The presence of entanglement can again be verified by measuring the conditional flux of anti-Stokes scattered photons \eqref{measure}, which now contains pairwise cross-terms $I(t) \propto \sum_j \langle a_j^\dagger a_j\rangle_t + 2 \sum_{i \neq j} {\rm Re}(\langle a_i^\dagger a_j \rangle_t)$. The latter oscillate with different effective mechanical detunings $\delta \omega_{\rm eff}^{ij} = \omega_i + \delta \omega_i^{\rm opt}- \omega_j -\delta \omega_j^{\rm opt}$ and decay with the respective decoherence rates $\gamma_{ij} = (\gamma_i + \gamma_j)/2$. For classically correlated states, the amplitude of the flux oscillations is bounded by $|\langle a_i^\dagger a_j \rangle_t| \leq \sqrt{\langle a_i^\dagger a_i a_j^\dagger a_j \rangle_t}$, so that observing a violation of this bound verifies the presence of entanglement in the state $\rho'$.

{\it Conclusion---} We demonstrated how entanglement between two dielectric nanospheres trapped in a single cavity mode can be created and read-out. This scheme is particularly suited for the setup of coherent scattering, where the tweezer is phase-coherently coupled to the cavity mode. By quantitatively   assessing  the expected signature in the presence of photon scattering decoherence, we argued that the observation of non-classical correlations between two nanoscale particles is an achievable goal in the foreseeable future.

We note that other entanglement schemes are conceivable for more conventional levitated setups, where the cavity is externally pumped \cite{niedenzu2012, abdi2015entangling}. While the presented protocol can be easily adapted to such a situation, the advantage of the coherent scattering scenario is that the cavity is empty in the mechanical groundstate, which substantially increases the signal-to-noise ratio and avoids laser phase noise \cite{delic2019}. Finally, the here proposed method may also serve to probe quantum correlations between different degrees of freedom of a single particle, such as the center-of-mass motion and the rotation of a nanorotor \cite{stickler2016a,kuhn2017a}.

{\it Acknowledgments} We thank Markus Aspelmeyer, Jack Clarke and Uro\v s Deli\' c for helpful discussions. B.A.S.\ acknowledges financial support by the European Union under the
Marie Sklodowska-Curie Grant Agreement 844010.

\end{document}